\documentclass[12pt,a4paper]{article}
\usepackage{amssymb}
\usepackage{amsmath}
\usepackage{amsfonts}
\usepackage{gensymb}
\usepackage{tabularx}
\usepackage{subcaption}
\usepackage{float,graphicx} 
\usepackage{natbib}
\usepackage{todonotes}
\usepackage{array,float,booktabs}
\usepackage[printonlyused]{acronym}
\RequirePackage[colorlinks,citecolor=black,urlcolor=blue]{hyperref}
\usepackage{url}

\DeclareMathOperator*{\argmin}{arg\,min}

\begin{document}

\def\spacingset#1{\renewcommand{\baselinestretch}%
{#1}\small\normalsize} \spacingset{1}	

\title{\bf Knots and their effect on the tensile strength of lumber: a case study}
\author{Shuxian Fan \\
    \footnotesize{Department of Statistics, University of Washington, Seattle WA, USA}\\
    Samuel W. K. Wong\footnote{Author for correspondence: samuel.wong@uwaterloo.ca} \\
    \footnotesize{Department of Statistics and Actuarial Science, University of Waterloo, Waterloo ON, Canada} \\
    James V. Zidek \\
    \footnotesize{Department of Statistics, University of British Columbia, Vancouver BC, Canada}
}
\maketitle

\bigskip
\begin{abstract}
When assessing the strength of sawn lumber for use in engineering applications, the sizes and locations of knots are an important consideration. Knots are the most common visual characteristics of lumber, that result from the growth of tree branches. Large individual knots, as well as clusters of distinct knots, are known to have strength-reducing effects. However, industry grading rules that govern knots are informed by subjective judgment to some extent, particularly the spatial interaction of knots and their relationship with lumber strength.  This case study reports the results of an experiment that investigated and modelled the strength-reducing effects of knots on a sample of Douglas Fir lumber. Experimental data were obtained by taking scans of lumber surfaces and applying tensile strength testing. The modelling approach presented incorporates all relevant knot information in a Bayesian framework, thereby contributing a more refined way of managing the quality of manufactured lumber.
\end{abstract}

\noindent%
{\it Keywords:} Bayesian predictive model; wood products; structural lumber testing; visual stress grading; strength-reducing characteristics; quality control
\vfill

\newpage
\spacingset{1} %DON'T change the spacing!

\section{Problem Description}\label{sec:probdesc}

\subsubsection*{The bench-scale stage of development}
This case study investigates the deleterious effects on sawn lumber strength due to the presence of \textit{knots}, which are the portions of branches or limbs cut through by a saw when lumber is produced from a log. More precisely, its novelty lies in its introduction of a new Bayesian approach for the construction of models to quantify those effects and predict strength along a specimen. It demonstrates, via a relatively small sample of Douglas Fir lumber, how that process of model development could take place. Details of the experiment and data are presented in Section \ref{sec:data}.

Overall, this case study falls within the bench-scale stage \citep{battaini2000bench} of process development, with its success on a small, inexpensive sample being used to establish a proof-of-concept. While our solution to the strength prediction problem works well in this case, the specific Bayesian model we developed is not unique and would require adaptation for other species, sizes and lengths. That leads in the concluding section to the recommendation that development should proceed to the pilot stage with a much larger sample and further model refinements. Success at that stage would lead to deciding whether a global (i.e., common across all producing facilities) or a mill-specific (i.e., unique to each producing facility) approach should be developed. The design of a more complete testing program can then follow to show the benefits and provide the impetus to establish the method as an acceptable standard practice. A standard practice would enable the method to be applied more broadly to different regions and species groups, and ultimately adopted as an industry policy for grading lumber. Other related applications include using the method to understand how the random distribution of knots within pieces of lumber impact the performance of assemblies made up of multiple pieces of lumber, or how closely-spaced small knots interact to produce a similar strength-reducing effect as a larger single knot.

\subsubsection*{Process capability analysis}
Due to the innate variability of lumber, methods for assessing its strength qualities are crucial to reduce consumer risk and ensure the safety of wood-based structures. The use of lumber as a construction material predates written history \citep{tegel2012early}. In Canada and the United States, its production grew rapidly around the turn of the 20th century \citep{reynolds1923lumber}, establishing lumber as a major industry. More recently, the interest in wood as a sustainable material has led to the rise of mass timber structures \citep{harte2017mass}.

Like any other manufactured product, lumber producers must monitor and ensure its quality. Although quality could be defined by a variety of criteria, the importance of any given criterion will be determined by consumers according to their intended application. For example, wane, which is the presence of bark, may devalue a piece of lumber where aesthetics are important. But in construction and load-bearing applications, strength is the most important measure of lumber quality. The need to specify quality at the time of manufacture leads to a trade-off between producer risk (e.g., underestimating the lumber's strength and value) and consumer risk (e.g., selecting lumber that fails to meet the strength requirements when placed in service). The idea of balancing these risks was recognized at least as far back as 1924, when rules were first established for sorting lumber into grades \citep{lumbergrades}.

The earliest grading rules were based on establishing several strength levels, expressed as a percentage of the ``clear'' or defect-free strength of the wood species of interest. Under each strength level, a cross-section of common strength-reducing visual characteristics (e.g., a knot at the edge of the wide face) and their maximum permissible size would be given. These grade rules, however, varied from region to region, even for the same wood species. Beginning in 1970, grading rules were harmonized so that North American structural lumber were all graded to the same rules. This then led to a large testing program in North America involving 70,000 pieces of sawn lumber, sampled from numerous mills in Canada and the United States \citep[e.g., see][for a review]{zidek2018statistical} to develop basic engineering properties, or design values (DVs), from tests on representative samples of production defined by the grading rules. Based on the test data, characteristic strength values (e.g., 5th percentile of the strength distribution) were estimated for the different grades, from which DVs were subsequently derived. Engineers following standard structural design equations could then determine which size, grade, and species of lumber to specify for each component in the wood structure, so that the requirements of the occupant and building regulator are both met.  This uniform system of assigning DVs underlies lumber manufacturing as we know it today.  

The famous so-called ``in-grade'' experiment described above could be thought of as a process capability analysis, in the terminology of modern quality control theory. Like any other process, it was anticipated that the DVs might decline due to assignable causes such as climate change, which could trigger, for example, an insect infestation attacking one wood species, or severe fire, ice or wind storms across a wide area to impact the wood harvest from a region. As a result, monitoring programs were established for North American lumber to detect changes to strength properties within grades \citep{kretschmann1999monitoring}. If significant changes were to occur, then production policies to maintain DVs, or updates to DVs or grading rules would be needed. Since visual characteristics are a key component of grading rules, understanding their relationship with strength is an important research problem.

The use of visual characteristics to predict lumber strength is well-established in the literature. For example, in so-called weak zone models, each piece of dimensional lumber is conceptualized as a composite of clear wood sections connected by weak sections, where the location of lumber defects can vary randomly~\citep{kandler2015stochastic, garcia2017deflections}. In the lumber tensile strength models proposed by~\citet{taylor1992comparing}, a piece of lumber is similarly considered to be comprised of smaller contiguous segments.  Regression methods have also been used to predict lumber strength based on associated characteristics, such as knot area ratios and the maximum diameters of the estimated strength-reducing knots \citep{divos1997lumber, francca2018modeling}. The efficacy of applying modern statistical methods for grading design purposes has also been established, as in a recent study by \citet{wong2016quantifying}. That work proposed a Bayesian approach for constructing a coherent hierarchical framework for modelling lumber strength, based on the characteristics recorded from the grading process and the uncertainties involved. Visual characteristics were recorded in accordance with grading rules, which included the knot deemed to be the most strength-reducing, the presence or absence of shake (a separation of lumber along its grain), and a miscellaneous category to cover other defects. 

Knots, formed from branches or limbs during the growth of the tree, constitute the most important strength-reducing characteristic of dimensional lumber~\citep{foley2003modeling}. Sawing practices result in \textit{knot faces} visible on lumber surfaces as dark elliptical cross-sections that cause fiber distortion. Clusters of knots, as well as other knots in close proximity, are known to have combined effects in reducing the strength of lumber~\citep{oh2010use,fink2014model}. However, grading rules only require the details for a single knot (or knot cluster) to be recorded, and selecting that maximum strength-reducing knot requires some subjective judgment. Due to these grading limitations, the complete spatial arrangement of knots has not been used in previous strength prediction models for lumber specimens, to the best of our knowledge. Therefore, the spatial interaction of knots and their relationship with strength properties requires further study and model development.

While knots appear on the surfaces of sawn lumber as elliptical shapes, these knot faces are only cross-sections of their underlying three-dimensional (3-D) structure. The idea of exploiting 3-D knot information to inform the strength of lumber was shown to be promising in \citet{lukacevic2015discussion}. To infer 3-D structure from the knot faces, a procedure known as \textit{knot matching} can be applied to determine the likely correspondence between visible ellipses and tree branches \citep{jun2019sequential}. This then allows the 3-D structure and volume of each knot to be reconstructed, based on the knot faces that are matched together. We leverage these recent developments in this case study so that the strength-reducing effects of knots can be modelled as accurately as possible.

\subsubsection*{Summary of case study}
Given the above considerations, this case study proposes a new approach for modelling the tensile strength of lumber, based on the observed spatial arrangement of knots and their inferred 3-D volumes. Rather than obtaining data from the grading process \citep[e.g., as in][]{wong2016quantifying}, this study extracts the full knot information directly from surface scans of specimens. We focus on tensile strength, which is measured by a destructive test that pulls longitudinally on the two ends of a piece of lumber until it breaks. Our modelling approach is designed to describe how the joint strength-reducing effects of knots vary longitudinally for individual lumber specimens, which leads to a hierarchical data generating process involving a vector of latent strength variables.
	
We adopt a Bayesian framework for inference, which provides a coherent framework to handle the modelling complexities along with Markov chain Monte Carlo (MCMC) techniques to sample from the posterior distribution. Predictions then account for estimation uncertainty via the more readily interpretable credible interval. With the goal of larger-scale implementation, the Bayesian paradigm also has the benefit of allowing the posterior obtained at any stage in model development to be used as prior engineering knowledge for input in the next stage of development. So the choice of prior in a bench-scale quality control setting, like the present case study, is not critical and might simply allow the posterior to be primarily informed by the data. Overall, we demonstrate the efficacy of the proposed approach via simulation studies and an analysis of experimental data collected from a sample of Douglas Fir specimens.

The remainder of the case study is laid out as follows.  A description of the Douglas Fir bench-scale experimental data that underlie our study is provided in Section~\ref{sec:data}. Section~\ref{sec:model} proposes our Bayesian approach for lumber tensile strength and its inference. A simulation study (Section \ref{sec:simstudy}) shows that our proposed model and approach leads to well-behaved inferential results. The analysis results for the Douglas Fir experimental data appear in Section~\ref{sec:application}. The predictive performance of the fitted model is also investigated and compared to baseline regression models via cross-validation. Interpretations of the estimated strength-reducing effects of knots are discussed. We conclude the case study in Section~\ref{sec:discussion} and provide recommendations to practitioners for future work.

\section{Data Description: Tensile Strength of Douglas Fir} \label{sec:data}

The modelling approach proposed in this case study is applied on experimental data from a sample of Douglas Fir (DF) lumber specimens. The sample specimens were sourced from the United States and delivered to the FPInnovations industrial research laboratory in Vancouver, Canada in late 2018. The wood species of the specimens was DF, which is widely used for making sawn lumber thanks to its strong and stable wood quality \citep{vikram2011stiffness}. The grade of the specimens was L3, according to NLGA standards \citep{nlga2017}. Following delivery and conditioning of the specimens, demands on equipment and technical staff led to some delays, and data collection began in May 2019. While the full sample consisted of around 300 specimens in total, the experiment was halted prematurely due to COVID-19 and the subsequent laboratory closure. A total of 113 specimens had completed testing by the end of 2019, which comprises the dataset available for this study.

The data for each specimen consists of knot information and strength measurements, obtained by the procedures described in the remainder of this section. No missing or corrupted values were present. Pre-processing of the raw photographs to facilitate extraction of knot information is described in our previous work \citep{pan2021ellipse}. Work on the present model development began in early 2020 and represents the first analysis of this new knot and strength dataset.

\subsubsection*{Preparation of lumber specimens}

Our DF specimens were 12-ft long nominal 2 by 6 boards (actual dimensions  $1.5'' \times 5.5'' \times 12'$). A visual examination of the specimens indicated that the lumber was of high quality, with the number of knots per specimen ranging from zero to eight (only knots of meaningful size are counted, i.e., 1/10th of the specimen width or larger). Following ASTM standards for testing lumber \citep{astmd198}, the specimens underwent a conditioning process (e.g., to stabilize moisture content) prior to testing, to ensure the reliability of the subsequent strength measurements.

To photograph the lumber surfaces, each specimen was fed through a scanner equipped with rollers and two sets of digital cameras; a diagram of the setup is shown in the Appendix of \citet{pan2021ellipse}. The knot faces on the photographs were then identified and annotated with the assistance of a team of undergraduate students. Next, the knot matching algorithm of \citet{jun2019sequential} was run to determine the most likely correspondence between knot faces and tree branches. The 3-D structure of a knot was then reconstructed as the convex hull containing its knot faces that were matched together, from which the volume displaced by the knot could be calculated. Finally, the stiffness (or \textit{modulus of elasticity}, MOE) of each specimen was measured via the transverse vibration method, according to the requirements of ASTM Standard Test Method D6874 \citep{astmvib}.

\subsubsection*{Lumber strength testing}

\begin{figure}[ht]
\centering
\includegraphics[width=\textwidth]{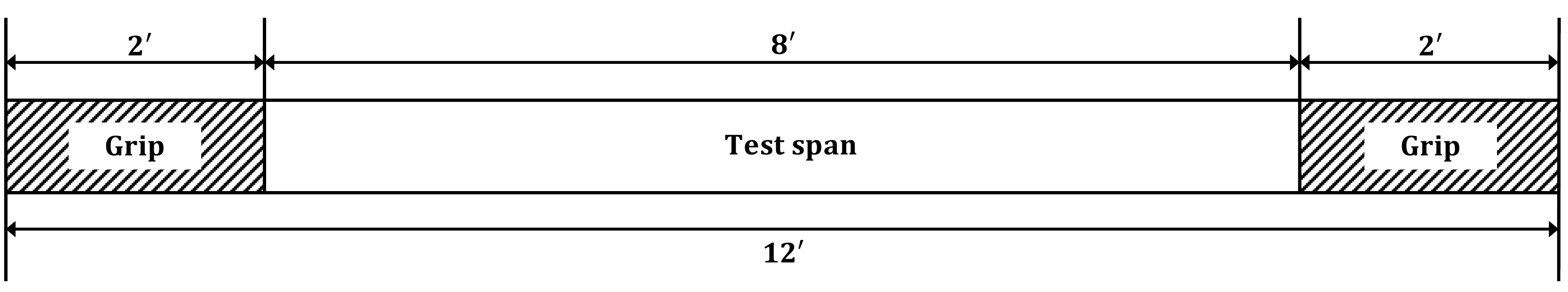}
\caption{Schematic diagram of the tensile strength test setup. Note the $8$-foot test span between the grips at each end.}
\label{fig:test_diagram}
\end{figure}

The tensile test machine, which meets the requirements of ASTM Standard Test Method D4761 \citep{astmtensile}, operates by pulling longitudinally on the two ends of the specimen; a schematic is shown in Figure~\ref{fig:test_diagram}. The machine has clamps that grip the specimen firmly at its two ends using a grip length of 2-ft, so with 12-ft specimens this results in an 8-ft test span between the grips. The grip is set to ensure that the tensile forces are axial can be assumed to be uniform along the specimen length between the inside edges of the grips. After the test begins, the tensile load applied (in lbs) is increased linearly at a constant rate and monitored via two load cells. The test is destructive in that all specimens are loaded until failure; the rate is set such that the average time taken to break a specimen is approximately 1 minute. The ultimate tensile strength (UTS) is calculated from the maximum load applied to the specimen immediately before it breaks, by converting to the equivalent pounds per square inch (psi). 

\begin{figure}[!ht]
\centering
\includegraphics[width=1\textwidth]{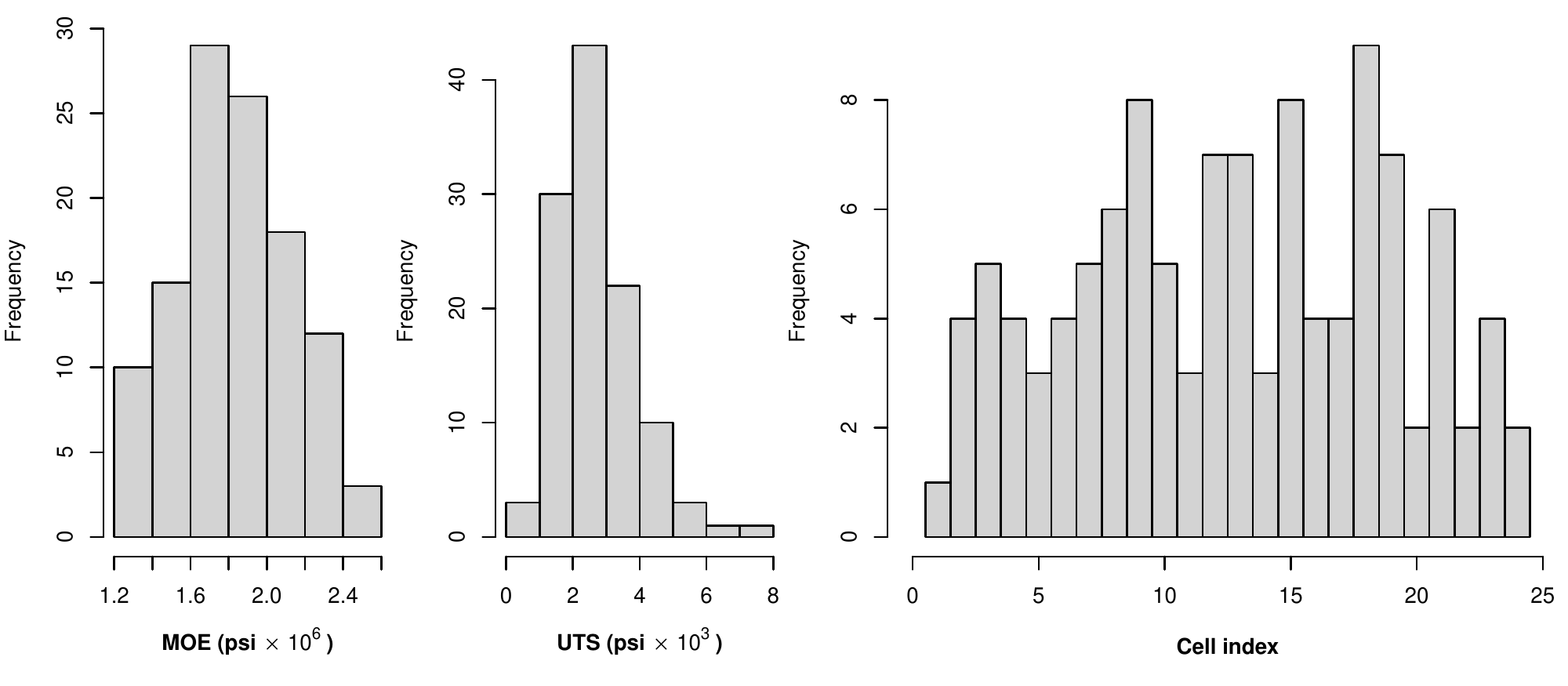}
\caption{Histograms of the MOE (psi$\times 10^6$), UTS (psi$\times 10^3$), and the cell indices of the failure locations (1--24), for the sample of 113 Douglas Fir specimens used in this study.}
\label{fig:vibE_uts_hist}
\end{figure}

After the tensile test is complete, the location of failure on the specimen must be determined by visual examination; this is defined by the location where the specimen first begins to fracture. Due to the setup of the grips, the specimen will fail at a longitudinal position between $24''$ and $120''$ from one end of the specimen. There is often some uncertainty in determining the exact location, because a fracture can be several inches long. Thus, we use a representation in Figure \ref{fig:vibE_uts_hist} that records a discretized cell index for the failure location, with each cell representing $4''$ of the test span between the grips. This $4''$ cell size represents the resolution at which most failure locations can be confidently determined, and yields a total of 24 cells for the 8-ft test span.

The measurements described in this section are summarized for the 113 specimens via the histograms in Figure \ref{fig:vibE_uts_hist}. The panels depict histograms of the MOE (psi$\times 10^6$), UTS (psi$\times 10^3$), and the cell indices of the failure locations (1--24). The empirical distributions of MOE and UTS are approximately symmetric, with UTS having a slight right skew. However, their relative spreads appear to be quite different, which may be summarized using the coefficient of variation (CV). MOE and UTS have CVs of $0.17$ and $0.45$ respectively, 
indicating that the latter is much more variable around its mean than the former. Note also that cell indices, recorded for the failure locations in our sample, are distributed relatively uniformly across the test span.

\section{Model for Tensile Strength}\label{sec:model}

\subsection{Model Conceptualization} 
\label{subsec:model-concept}

In general, the tensile strength of a lumber specimen varies continuously along its length, although in practice strength can only be observed at the discrete locations where failure occurs \citep{lam1991variation}. The combination of its visual characteristics, including knots, may also be considered as realizations of stochastic processes on a continuous spatial domain (i.e., the length and width of the specimen). The second may then be treated as a predictor of the first, while recognizing that other unobservable features can also contribute to its length-wise strength variability. Within a Bayesian framework, this leads us to the construction of a hierarchical model for tensile strength that leverages ideas from spatial statistics to describe the data generating process. One of the goals of this case study is to illustrate how that task of model development takes place, on the basis of the available data.
	
For a given lumber specimen in our dataset, we observe a collection of knots with various sizes and locations, and this section develops a framework for modelling the impact of the knots on tensile strength. To provide a realistic characterization of strength, the model should account for the recorded features of all strength-reducing knots, and other sources of uncertainty inherent to the specimen itself. To do so, we adopt a hierarchical structure that nests a process model ($X$) within a measurement model ($Y$). 

\begin{figure}[!htp]
\centering
\includegraphics[width=\textwidth]{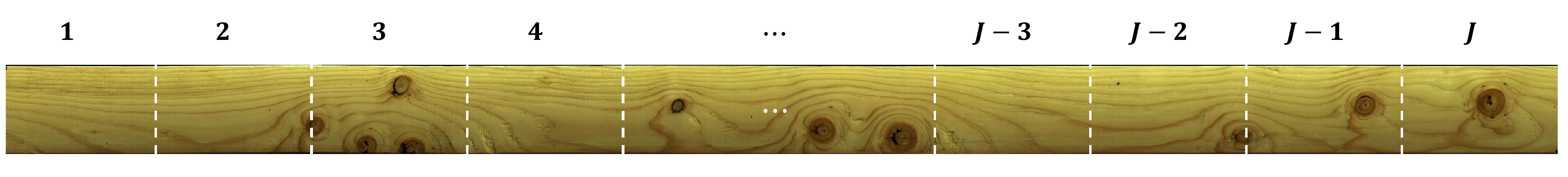}
\caption{Partitioning a lumber specimen into $J$ cells, using the surface of a sample specimen to illustrate. The dark elliptical shapes are the visible knot faces, and the cell boundaries are marked by the dotted lines. For example, there are four knot faces that are distinctly visible within cell 3.}
\label{fig:illustration}
\end{figure}

The process model is conceptualized by considering a ``clear'' specimen with no knots, and the measurement model then adjusts for the strength-reducing effects of knots. The idea is to then partition the specimen longitudinally into adjacent rectangular cells as a discrete approximation of the continuous variation, similar to the segmenting done in previous studies  \citep{taylor1992comparing}. Figure \ref{fig:illustration} illustrates such a partitioning with $J$ cells, where the knot faces from the surface of a sample specimen are visible. In practice, the number of cells $J$ could be chosen to reflect the highest resolution at which failure locations can be accurately determined from a tensile strength test; that is, after testing we can  observe the failure to have occurred in cell $j \in \{1,2, \ldots, J\}$. Then, specifying the process model over each cell provides a discretized representation of how strength varies along the piece of lumber.

\subsubsection*{Clear specimens}
We define a ``clear'' specimen to be one that has no knots. Consequently, the within-specimen strength variability of a clear specimen arises from other uncertainties associated with destructive strength testing and potentially other strength-reducing characteristics apart from knots.  We then assume the strength varies randomly along the length of a lumber specimen, with strengths of adjacent cells being autocorrelated. Evidence of this autocorrelation has been previously reported in the tensile experiment of \citet{lam1991variation}, where the unbroken portions of each specimen were re-used for further testing to obtain multiple strength measurements per specimen.

Let the strength of the cells in a ``clear'' specimen be represented by the vector $\pmb{X} =   \{X_1,\dots, X_J\} \in \mathbb{R}^J$.  Note that in practice, $\pmb{X}$ will be an entirely latent vector for any specimen that has knots. Thus, for simplicity we use an $AR(1)$ process of the form
\begin{equation}\label{eq:ar}
\nonumber 
    X_j = \alpha + \rho X_{j-1}+\epsilon_j, \hspace{0.5cm} 0<\rho<1,\hspace{0.5cm} \epsilon_j \stackrel{iid}\sim N(0, \sigma^2),
    \hspace{0.5cm} j=2,\ldots, J
\end{equation}
to model the autocorrelation among the $X_j$'s, where the mean is $\mu = E[X_j] = \alpha/(1-\rho)$ for $j=1, \ldots, J$ and $X_1 \sim N(\mu, \sigma^2/(1-\rho^2))$. Additional global or local covariates can be included in the framework by replacing $\alpha$ with a regression component $\pmb{\eta}^T\pmb{v}_j$, i.e., 
\begin{equation}\label{eq:covariate}
    E[X_j | X_{j-1} = x_{j-1}, \pmb{v}_j] = \rho x_{j-1} +  \pmb{\eta}^T\pmb{v}_j\;,
\end{equation}
where $\pmb{v}_j$ denotes the covariate vector for cell $j$ and $\pmb{\eta}$ is a vector of regression coefficients. If only global covariates (i.e., those common among all cells of the specimen) are used, then the $AR(1)$ process is stationary. Covariates could include any features of the specimen that can be measured non-destructively (i.e., without breaking the specimen) that are associated with strength. The appropriateness of the $AR(1)$ assumption can be  subsequently assessed via posterior predictive checks on the fitted model.

\subsubsection*{Adjusting for knots}

Many previous studies have demonstrated the significant impact of knots, and the resulting fiber deviations, on the strength properties of dimensional lumber \citep{castera1996prevision,cramer2007model,hietaniemi2011camera}. As discussed in \citet{foley2003modeling}, knots most likely cause the applied stress to be redistributed to a smaller cross-sectional area of clear wood, thus reducing tensile strength. In particular, knots that straddle the edge of a piece (or \textit{edge knots} for short) are considered to have larger strength-reducing effects compared to other types of knots~\citep{courchene1998effect}.  Therefore, it is important to consider both the overall spatial arrangement of knots and the placement of each individual knot within a specimen. For example, in the specimen shown in Figure \ref{fig:illustration}, the bottom two knot faces visible in cell 3 are edge knots. 

The number and location of knots on a specimen are random and might be conceptualized as a realization of a spatial point process. Here, we simply take their observed spatial arrangement as given, after extracting all relevant knot information from scans of the lumber surfaces. To be more precise, suppose a specimen has $K$ knots, and let $Z_k$ denote the intrinsic strength-reducing effect of the $k^{\rm th}$ knot, $k=1, \ldots, K$, which in our case study is taken to be the 3-D volume displaced by the knot \citep{lukacevic2015discussion}. Then, we define the measurement model $\pmb{Y}$ to be the process model adjusted by the joint strength-reducing effects of all $K$ knots. For a particular cell $j$, its adjusted strength $Y_j$ will depend on knots within the cell, together with contributions from knots in other cells albeit to a lesser extent. Intuitively, the impact of a knot will decrease according to its relative distance from the cell.

Given the above considerations, we model $Y_j$ with the form
$$
Y_j = X_j - \sum_{k = 1}^{K} \gamma_{(k)} h(d_{jk}) Z_{k},
$$
where $d_{jk}$ is the distance between the centroids of knot $k$ and cell $j$, and $h(d_{jk})$ is a chosen decreasing function of $d_{jk}$. The coefficient $\gamma_{(k)}$ then scales these distance-weighted knot effects to a corresponding reduction in strength for the cell, defined according to
\[\gamma_{(k)} = \begin{cases} 
      \gamma_0 & \text{if the $k^{th}$ knot is not an edge knot} \\
      \gamma_1 & \text{if the $k^{th}$ knot is an edge knot}.
\end{cases}\]

\subsubsection*{Strength measurement}

Tensile strength testing applies loads uniformly along the length of the specimen, so that the specimen is expected to break at its weakest location. In our discretized representation of strength along the specimen, the observed strength will thus be that of the weakest cell among the $J$ cells, where the index of the cell corresponding to the location of failure is also observed. Specifically, the observed strength measurement $Y_{obs}$ for the specimen is $Y_{obs} = Y_m$, where $m=\argmin_j\{Y_j : j= 1, \ldots, J \}$. This construction induces truncated distributions for the strengths in other cells, since $Y_j > Y_m$ for $j \ne m$.

\subsection{Fitting Model Parameters}\label{subsec:inference}

We adopt a Bayesian approach to infer the model parameters in the preceding framework, given the observed data from a sample of $n$ specimens. This allows us to incorporate relevant prior information into the model and to obtain predictive distributions for the strengths of future specimens that coherently account for the posterior uncertainties of the parameters.

Considering specimen $i$ of the sample, $i = 1, \hdots, n$, let $K_i \ge 0 $ be the number of knots in the specimen and $\pmb{S}_{i}=[\pmb{s}_1,\hdots,\pmb{s}_{K_i}]$ their corresponding knot centroids. Denote the centroids of the $J$ cells in the partitioning by $[\pmb{c}_1,\pmb{c}_2,\hdots,\pmb{c}_J]$, then define the distance matrix $\pmb{D}_i \in \mathbb{R}^{J \times K_i}$ as 
\[	\pmb{D}_i  = [\pmb{d}_1,\pmb{d}_2,...,\pmb{d}_{J-1}, \pmb{d}_J]^T\;,\]
where $\pmb{d}_j = ||\pmb{S}_{i} - \pmb{c}_j||_2 \in \mathbb{R}^{K_i}$ is the vector of Euclidean distances between the $j^{\rm th}$ cell and each knot.  In this case study we use a simple exponential decay \citep[e.g.,][]{biggeri1996air} for the distance function,
$h(d) = \exp\{{-\beta}d\}\mathbb{I}(d \leq d_{\max})$ where $\beta$ is a  parameter to be estimated and $d_{\max}$ is the maximum distance at which a knot can have an impact on strength. We also considered other commonly used distance-decaying functions to model the effect of point sources, such as a power of distance ($d^{-\beta}$) and Gaussian decay ($\exp\{-\beta d^2\}$) \citep{bithell1995choice, diggle1990point}, and found that exponential decay provided the best empirical performance in this case. Then the weight matrix $\pmb{W}_i\in \mathbb{R}^{J \times K_i}$ is obtained by applying the function $h(\cdot)$ to $\pmb{D}_i$ element-wise. Finally, let $\pmb{Z}_i \in \mathbb{R}^{K_i\times 1}$ denote the intrinsic strength-reducing effects of the $K_i$ knots and $\pmb{E}_i \in \{0,1\}^{K_i \times 1}$ be a vector of indicators, with 1 denoting an edge knot. All of $K_i$, $\pmb{D}_i$, $\pmb{Z}_i$, and $\pmb{E}_i$ will be observed for the specimen.

Next we turn to the strength model, and let $\pmb{X}_i \in \mathbb{R}^{J\times 1}$ denote the strengths of the $J$ cells when specimen $i$ is clear of knots, assumed to follow an $AR(1)$ process and have a normal distribution as presented in Section \ref{subsec:model-concept}. Taking the setup in Equation (\ref{eq:covariate}) where $\pmb{X}_i$ also depends on covariates, let $\pmb{v}_{ij}$ denote the relevant covariates for cell $j$, and $\pmb{\eta}$ the vector of coefficients. Then  $\pmb{Y}_i$ is the vector of cell strengths adjusted for knots and given by
\begin{equation}\label{eq:ymat}
\pmb{Y}_i =  \pmb{X}_i  -  \pmb{W}_i \times \left( \{  \gamma_0 (\pmb{1}_{K_i} - \pmb{E}_i) + \gamma_1 \pmb{E}_i \} \circ \pmb{Z}_i \right), 
\end{equation}
where $\pmb{1}_{K_i}$ is a column vector of $K_i$ ones and $\circ$ is the element-wise product. After the tensile test, denote the strength measurement observed for the specimen as $Y_{obs, i}$, which is defined by the minimum of the underlying adjusted strengths, i.e., $Y_{obs, i} = \min \{ \pmb{Y}_i \}$ and let $m_i$ denote the corresponding cell index of the failure location.

Then the model parameters to be inferred from the sample are in vectorized form,  $\pmb{\theta} = [\pmb{\eta}, \rho, \sigma^2,\beta,\gamma_0, \gamma_1]$, which has posterior distribution given by
\begin{align}
p(\pmb{\theta} |  \left\{Y_{obs, i}, m_i, K_i, \pmb{D}_i, \pmb{Z}_i, \pmb{E}_i \right\}_{i=1}^n)  \propto \pi(\pmb{\theta} ) \prod_{i=1}^n p( Y_{obs, i}, m_i | K_i, \pmb{D}_i, \pmb{Z}_i, \pmb{E}_i, \pmb{\theta}) \nonumber \\ 
\propto  \pi(\pmb{\theta} ) \prod_{i=1}^n \int_A p( \pmb{Y}_i | K_i, \pmb{D}_i, \pmb{Z}_i, \pmb{E}_i, \pmb{\theta})\, d \pmb{Y}_{-obs,i}
\end{align}
where $\pi(\pmb{\theta})$ denotes the joint prior distribution of the parameters, $\pmb{Y}_{-obs,i}$ is $\pmb{Y}_i$ with element $m_i$ deleted, and the integral is taken over $A = ({Y_{obs,i}},{\infty})^{J-1}$. The term $p( \pmb{Y}_i  | K_i, \pmb{D}_i, \pmb{Z}_i, \pmb{E}_i, \pmb{\theta})$ can be expressed using normal distributions according to the setup for $\pmb{X}_i$ and Equation (\ref{eq:ymat}).

This posterior is analytically intractable, so we adopt %Markov chain Monte Carlo (MCMC)
MCMC techniques for inference.  To facilitate computation, we augment the posterior by including the latent variables $\pmb{Y}_{-obs,i}$, that is, we draw MCMC samples for $\pmb{\theta}$ and $\pmb{Y}_{-obs,i}$ jointly from
\begin{align*}
p(\pmb{\theta}, \left\{ \pmb{Y}_{-obs,i} \right\}_{i=1}^n|  \left\{Y_{obs, i}, m_i, K_i, \pmb{D}_i, \pmb{Z}_i, \pmb{E}_i \right\}_{i=1}^n)   \propto  \pi(\pmb{\theta} ) \prod_{i=1}^n p( \pmb{Y}_i | K_i, \pmb{D}_i, \pmb{Z}_i, \pmb{E}_i, \pmb{\theta})\cdot I_A(\pmb{Y}_{-obs,i})
\end{align*}
where $I_A(\pmb{Y}_{-obs,i})$ is the usual indicator function that takes value 1 when $\pmb{Y}_{-obs,i} \in A$. We carry out the required MCMC sampling via Hamiltonian Monte Carlo (HMC) as implemented in Stan \citep{stan2018rstan} using the R statistical computing environment.

\subsection{Model validation} \label{sec:simstudy}

We first fit our proposed model to simulated data and examine the posterior distributions of the parameters obtained by Bayesian inference. To mimic the subsequent real data analysis, we set $J = 24$, $d_{max} = 96$ (inches), and take $\pmb{\eta} = \{\eta_0, \eta_1\}$ to be the regression coefficients in Equation (\ref{eq:covariate}) since each specimen in the real data has one covariate. We also choose values for the model parameters that resemble those found by a preliminary analysis of the real data: $\eta_0 = 3.0$, $\eta_1 = 1.5$, $\rho = 0.7$, $\sigma = 0.8$, $\beta = 0.5$, $\gamma_0 = 0.25$, $\gamma_1 = 0.15$. Then, we  simulate datasets of three different sample sizes:  $n = 120, 360, 720$ specimens.  The standard sample size in lumber testing is 360 specimens \citep{green1993investigation}, and 120 is close to the sample size of the available data in the present study.

We next detail the steps involved in generating a simulated dataset corresponding to the given parameters and values, independently for each specimen $i=1,2, \ldots n$:

\begin{enumerate}
	\item Generate the covariate $v_i \sim N(1.9, 0.25^2)$. As described in Section~\ref{sec:data}, this covariate is a measurement of the specimen's overall stiffness (MOE). We treat this as a global covariate common to all $J$ cells. The mean 1.9 and  standard deviation 0.25 of the normal distribution are chosen to resemble the empirical distribution of MOE (in units of psi $\times 10^6$) in the real data.
	
    \item Generate $\pmb{X}_i \in \mathbb{R}^{J\times 1}$, i.e., the strengths of each cell for a clear specimen, in successive order according to
	 \begin{equation}
	 	\begin{aligned}
	 	X_1 &\sim N(\eta_0 + \eta_1 v_i, \sigma^2/(1-\rho^2)) \;,\\
	 	X_2 &= (1-\rho)(\eta_0 + \eta_1 v_i) + \rho X_1 + \epsilon_2\;,\\
	 	\vdots\\
	 	X_J &= (1-\rho)(\eta_0 + \eta_1 v_i) + \rho X_{J-1} + \epsilon_{J}\;,
	 	\end{aligned}\label{eq:x-generative}
	 \end{equation}
	where $\epsilon_2, \ldots, \epsilon_J \stackrel{iid}\sim N(0, \sigma^2)$. 
	The parameters were chosen so that there would be a negligible probability of negative strength values, thus enabling the normal distribution to serve as a reasonable approximation in practice.
 
    \item Generate the number of knots $K_i$ and locations of the knot centroids $\pmb{S}_{i}$ from a homogeneous Poisson point process with constant intensity $\lambda = 0.01$ within the dimensions of the test span ($96 \times 5.5$ in$^2$). The intensity $\lambda$ is chosen to resemble the number of knots observed in the real data. We then compute the distance matrix $\pmb{D}_i\in \mathbb{R}^{J \times K_i}$ given $\pmb{S}_{i}$, and apply the decreasing function $h(\cdot)$ element-wise on $\pmb{D}_i$ to obtain the weight matrix $\pmb{W}_i\in \mathbb{R}^{J \times K_i}$.

    \item Generate the vector of indicators $\pmb{E}_i \in \{0,1\}^{K_i}$ to denote  whether each knot is an ``edge knot'' or not. Each element of $\pmb{E}_i$ is obtained as an independent Bernoulli($p_e$) draw, and the value of $p_e = 0.6$ is chosen based on the observed proportion of edge knots in the real data. 
    
    \item  Generate the vector of strength-reducing effects $\pmb{Z}_i \in \mathbb{R}^{K_i\times 1}$. Each element of $\pmb{Z}_i$ is obtained as an independent draw from a gamma distribution with shape parameter 2.0 and scale parameter 6.0, again informed by the knot volumes extracted from the real data.

    \item  Calculate $\pmb{Y}_i \in \mathbb{R}^{J\times 1}$ conditional on the outputs of simulation steps 1--5, using Equation (\ref{eq:ymat}).  We obtain the observed strength measurement $Y_{obs,i}$ as the minimum of $\pmb{Y}_i$ and record $m_i$ as the corresponding cell index where failure occurred.
\end{enumerate}

We simulate a dataset for each of the three aforementioned sample sizes of specimens (120, 360, and 720). To complete the model specification, we chose weakly informative priors for the parameters as follows:
\begin{itemize}
	\item $\eta_0 \sim N(0, 10)$, $\eta_1 \sim N(0, 10)$;
	\item $\rho \sim N(0.5,0.5^2)$ restricted to $0 < \rho < 1$;
	\item $\beta \sim N(0,1)$, $\gamma_0 \sim N(0,1)$, $\gamma_1 \sim N(0,1)$, each restricted to the positive half of the standard normal;
	\item $\sigma \sim Cauchy(0,5)$.
\end{itemize}
These priors follow the general recommendations of the Stan developers \citep[][https://github.com/stan-dev/stan/wiki/Prior-Choice-Recommendations]{gelman2017prior,stan2018rstan}, where the coefficients $\eta_0, \eta_1, \beta, \gamma_0, \gamma_1$ are given fairly vague normal priors (relative to their magnitude) and the scale parameter $\sigma$ is given a weak Cauchy prior. The model setup requires $\beta, \gamma_0, \gamma_1$ to be positive, and a positive correlation $\rho$ is consistent with past experiments \citep{lam1991variation}. While there is no unique choice for a weakly informative prior, a reasonable choice will allow the posterior to be primarily informed by the data, which the simulation results confirm. Moreover for future applications of the methodology, the principle of Bayesian updating means that today's posterior can become tomorrow's informative prior, namely a choice guided by the data already analyzed in this case study.

For each simulated dataset, we fit the proposed model using the methods described in Section \ref{subsec:inference}. We ran four parallel chains with 10000 HMC iterations each, and discarded the first 5000 iterations as burn-in. The posterior distributions of the parameters are summarized in Table~\ref{tab:sim_parameters} by displaying their 50\%, 2.5\%, and 97.5\% posterior quantiles. The fits seem reasonable, and it can be seen that the 95\% credible intervals (represented by the 2.5\% and 97.5\% posterior quantiles) all contain the true parameter values shown in the `Truth' column, which indicate that the priors do not strongly contribute to the posterior.  The widths of these credible intervals also become narrower as the sample size increases. These results indicate that we can expect to obtain useful parameter estimates with a sample size of around 120 specimens, while more precise credible intervals could be obtained with a standard test sample of 360 specimens. The additional precision gained from a further increase to 720 specimens is less pronounced. 

\begin{table}[ht]
\caption{Summaries of the posterior distributions of the parameters based on simulated datasets with different sample sizes of specimens ($n = 120$, $360$, $720$).  The true parameter values are shown in the `Truth' column. The 50\%, 2.5\% and 97.5\% posterior quantiles are shown for the parameters in each simulated dataset. The 95\% credible intervals all contain the true parameter values, and become narrower as the sample size increases. \strut}
	\centering
	\begin{tabular}{ccccccccccc}
		& & \multicolumn{9}{c}{Posterior quantiles} \\  \cmidrule(lr){3-11}
		& & \multicolumn{3}{c}{$n = 120$} &  \multicolumn{3}{c}{$n = 360$} &  \multicolumn{3}{c}{$n = 720$}\\ \cmidrule(lr){3-5}  \cmidrule(lr){6-8} \cmidrule(lr){9-11}
		Parameter & Truth & 50\% & 2.5\% & 97.5\% & 50\% & 2.5\% & 97.5\% & 50\% & 2.5\% & 97.5\% \\ 
		\hline
		$\eta_0$ & 3.00 & 3.82 & 2.68 & 4.96 & 3.24 & 2.52 & 3.97 & 3.16 & 2.66 & 3.66 \\ 
		$\eta_1$ & 1.50 & 1.12 & 0.61 & 1.64 & 1.30 & 0.97 & 1.62 & 1.46 & 1.24 & 1.69 \\ 
		$\rho$ & 0.70 & 0.64 & 0.26 & 0.83 & 0.74 & 0.61 & 0.83 & 0.67 & 0.56 & 0.75 \\ 
		$\sigma$ & 0.80 & 0.83 & 0.54 & 1.20 & 0.71 & 0.56 & 0.90 & 0.86 & 0.72 & 1.01 \\ 
		$\beta$ & 0.50 & 0.48 & 0.30 & 0.75 & 0.49 & 0.39 & 0.62 & 0.52 & 0.43 & 0.62 \\ 
		$\gamma_0$ & 0.25 & 0.23 & 0.13 & 0.48 & 0.24 & 0.17 & 0.35 & 0.26 & 0.20 & 0.35 \\ 
		$\gamma_1$ & 0.15 & 0.13 & 0.06 & 0.28 & 0.13 & 0.09 & 0.20 & 0.15 & 0.11 & 0.20 \\ 
		\hline
	\end{tabular}\label{tab:sim_parameters}
\end{table}

\section{Case Study: Analysis and Interpretation} \label{sec:application}

This section analyzes the experimental data presented in Section~\ref{sec:data} using the approach proposed in Section~\ref{sec:model}. Comparisons of predictive performance with simpler regression models are also considered. Lastly, interpretations of the estimated knot effects are described.

\subsection{Model fitting results}\label{subsec:modelfitting}

We fit the proposed model to the knot and tensile strength experimental data described in Section \ref{sec:data}, where we set $J = 24$ and $d_{max} = 96$in to correspond to the length of the test span. Since the MOE is measured for each specimen as a whole, we treat it as a global covariate that is common to all $J$ cells of a specimen and take $\pmb{\eta} = \{\eta_0, \eta_1\}$ to be the regression coefficients in Equation (\ref{eq:covariate}); i.e., as in the simulation study (Section \ref{sec:simstudy}) we have for an individual specimen $\pmb{\eta}^T\pmb{v}_j = \eta_0 + \eta_1 \cdot MOE$,  $j=1, \ldots, 24$. We also adopt the same priors as introduced in the simulation study.

We ran four parallel chains with 10000 HMC iterations each and discarded the first 5000 iterations as burn-in. Convergence was observed in the trace plots, and summaries of the posterior distributions for the parameters are reported in Table~\ref{tab:parameters}. The widths of the 95\% credible intervals (represented by the 2.5\% and 97.5\% quantiles) are comparable to those from the 120 specimen simulation study in Table \ref{tab:sim_parameters}.

\begin{table}[ht]
	\centering
\caption{Summary of the posterior distributions of the parameters, based on fitting the model to the Douglas Fir experimental data. The $50\%$, $2.5\%$, and $97.5\%$ posterior quantiles are shown for each parameter.}
	\begin{tabular}{cccc}
         & \multicolumn{3}{c}{Posterior quantiles} \\ \cmidrule(lr){2-4} 
 		Parameter  & 50\% & 2.5\% & 97.5\% \\ 
		\hline
		$\eta_0$ & 2.59 & 1.20 & 4.03 \\ 
		$\eta_1$ & 1.69 & 1.05 & 2.33 \\ 
		$\rho$ & 0.76 & 0.52 & 0.88 \\ 
		$\sigma$ & 0.94 & 0.61 & 1.44 \\ 
		$\beta$ & 0.40 & 0.25 & 0.64 \\ 
		$\gamma_0$ & 0.30 & 0.18 & 0.55 \\ 
		$\gamma_1$ & 0.12 & 0.07 & 0.23 \\ 
		\hline
	\end{tabular}\label{tab:parameters}
\end{table}

\begin{figure}[ht]
	\centering
	\includegraphics[width=\textwidth]{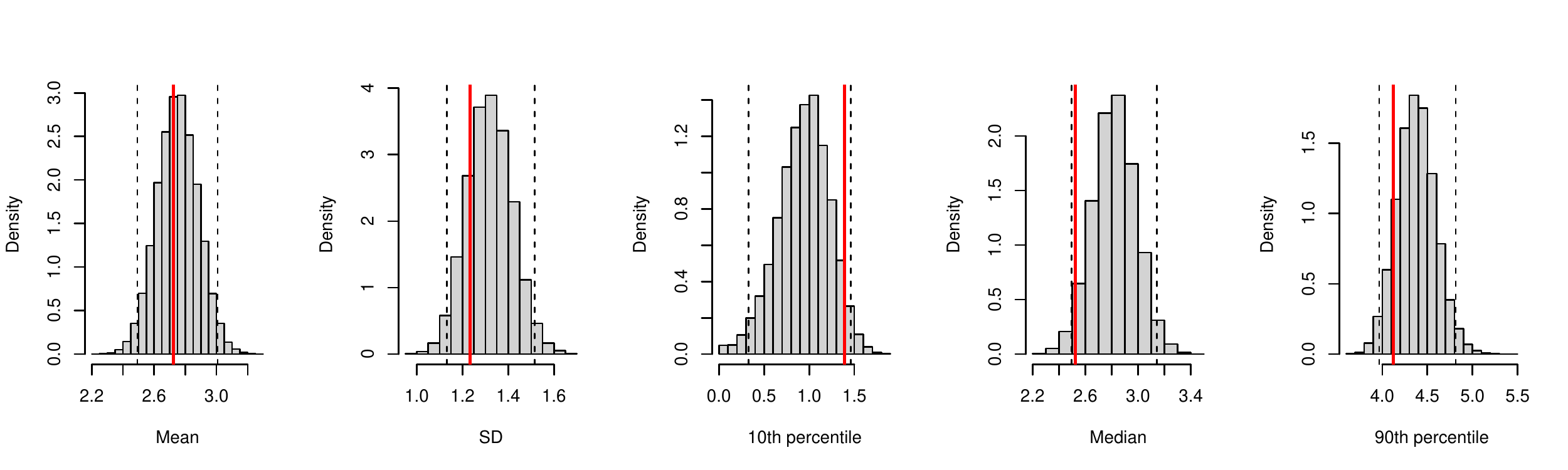}
	\caption{Posterior predictive checks by generating replicated tensile strength datasets from the model using the posterior distribution of the parameters. Test quantities include the mean, standard deviation, and percentiles (10th, 50th, and 90th) of UTS. The histograms show the predictive distributions of these quantities over the replicated datasets, with dotted vertical lines indicating the central 95\% credible intervals. The solid vertical line indicates the observed value from the real data.}
	\label{fig:postpred}
\end{figure}

Posterior predictive checks \citep[e.g.,][]{gelman2013bayesian} are used to assess whether the observed UTS measurements are adequately explained by the model. To do so, we generate a replicated dataset for UTS conditional on each MCMC draw of the posterior parameters and the observed knot information. These predictive datasets are summarized via five test quantities: mean, standard deviation, and percentiles (10th, 50th, and 90th). Histograms for each quantity are displayed in Figure \ref{fig:postpred}, which indicate that the observed value for each of these quantities in the real dataset lies within the central 95\% interval of the predictive distribution. The fit appears adequate for this case study, while recognizing that future applications of our methodology might require adaptations to some of the modelling assumptions, such as the normal distributions used here for $\pmb{X}$.

\subsection{Strength prediction}\label{subsec:model_pred_performance}

The fitted Bayesian model provides a posterior predictive distribution for lumber tensile strength, given the MOE and spatial arrangement of knots (i.e., as represented via $\pmb{W}_i,\pmb{Z}_i, \pmb{E}_i$ in the modelling framework) for a new specimen. An estimate of  this predictive strength distribution is obtained by taking each MCMC draw of the parameters, using Equation (\ref{eq:x-generative}) to generate $\pmb{X}_i$, then using Equation (\ref{eq:ymat}) to generate $\pmb{Y}_i$ and setting $\min\{\pmb{Y}_i\}$ to be the tensile strength.

To assess the efficacy of our modelling approach, we performed five-fold cross-validation to evaluate predictive performance. To provide a baseline for comparison, we consider our Bayesian model along with two basic regression models for UTS: 
\begin{equation}
\begin{aligned}
 \text{Regression Model 1}:   & \text{UTS} = \beta_0 + \beta_1\cdot \text{MOE} + \epsilon\;, \\
 \text{Regression Model 2}:   & \text{UTS} = \Tilde{\beta_0} + \Tilde{\beta_1}\cdot \text{MOE} + \Tilde{\beta_2}\cdot \max(\pmb{Z}_i)  + \Tilde{\epsilon}\;.
    \nonumber
\end{aligned}
\end{equation}
The first model only uses the MOE as a predictor, while the second model also includes the worst individual knot effect, defined via $\max(\pmb{Z}_i)$ in our notation. Inclusion of the worst knot mimics the setup of some previous regression models for lumber strength \citep[as described in the Problem Description, e.g.,][]{divos1997lumber}.

For our Bayesian model, we treat the mean of the posterior predictive distribution as the predicted strength for the validation specimens, and consider the 2.5\% and 97.5\% quantiles as a 95\% prediction interval.  The cross-validated prediction results are summarized for the three models in Table~\ref{tab:preds}. The mean predicted tensile strength for each model over the 113 specimens is close to the empirical mean UTS in the dataset of 2.72, which indicates that our Bayesian predictions have little or no bias in practice. As metrics to evaluate the predictive performance of each model, we compute the mean-squared prediction error (MSPE), mean absolute prediction error (MAPE), and mean length of the 95\% prediction intervals. Including the worst knot in the basic regression model leads to a marked improvement in MSPE compared to a regression on MOE alone (1.04 vs.~1.23), and our proposed Bayesian model that accounts for the entire spatial arrangement of knots yields a modest further improvement (MSPE $=0.96$). A similar pattern is seen when considering MAPE. The Bayesian model also provides the narrowest 95\% prediction intervals on average.

These results suggest the importance of knots and their spatial arrangement in lumber strength prediction, while recognizing that statistical significance will require a larger sample, as seen via the standard errors in Table \ref{tab:preds}. Intuitively, the advantage of the Bayesian model in predictive performance might be most evident in specimens with multiple larger knots, where their spatial arrangement can have a more important role. Indeed, for the 51 specimens in the sample having at least three knots with volume greater than 10\% of the cell size, the MSPEs were 1.27 for the Bayesian model and 1.50 for Regression Model 2. Finally, our results were quite robust to the choice of the number of cells $J$ used for computation: the MSPEs for $J=12$ and $J=48$ were 0.96 and 0.98 respectively, indicating that a range of reasonable choices for $J$ in our model can work well in practice.

\begin{table}[!ht]
	\centering
\caption{Summary metrics for the predictive performance of two linear regression models and the proposed Bayesian model, by applying five-fold cross-validation on the Douglas Fir dataset.  The mean tensile strength in the dataset is 2.72 and close to the mean predictions given by the three models under cross-validation. The mean-squared prediction error, mean absolute prediction error, and mean length of 95\% prediction interval are computed for each model, with standard errors in brackets.}
	\begin{tabular}{lccc}
		     & \multicolumn{3}{c}{Model} \\ \cmidrule(lr){2-4} 
		Metric & Regression 1 & Regression 2 & Bayesian spatial \\ \hline
		Mean prediction & 2.71 (0.06) & 2.71 (0.07) & 2.73 (0.08)\\
		Mean-squared prediction error & 1.23 (0.21) & 1.04 (0.18) & 0.96 (0.16) \\
		Mean absolute prediction error & 0.88 (0.02) & 0.80 (0.02) & 0.78 (0.04)\\
		Mean length of 95\% prediction interval & 4.37 (0.06) &  4.06  (0.06) & 3.94 (0.06)\\\hline
	\end{tabular} \label{tab:preds}
\end{table}

\subsection{Interpretation of Knot Effects}
By conceptually partitioning a specimen into cells, our fitted model helps provide insight into a knot's effect as a function of distance. The posterior distribution of the model parameter $\beta$ suggests that the spatial effect of a knot decays at a rate of approximately $e^{-0.40 d}$, where $d$ denotes the distance (in inches) between the knot location and a cell centroid.  (Note that the quantity $0.40$ has units in$^{-1}$ so that the exponential is unitless.) Thus, the estimated spatial effect decays to $< 1\%$ of the maximum at a 1-ft distance from the knot. Interestingly, \citet{barbosa2019effect} chose a fixed cutoff of 6-in for considering multiple knots to be in close proximity, which is qualitatively similar to our inferred result. They concluded via linear regression models that such knots work together to reduce strength, albeit in the context of bending and the Southern Pine species.

Moreover, our fitted model sheds light into the effect of a knot's location, via the parameters $\gamma_0$ and $\gamma_1$ that scale a knot's strength reduction depending on whether it is situated on the edge of a specimen.  Based on the $113$ specimens available, the 95\% credible intervals were $(0.18, 0.55)$ for an edge knot and $(0.07,0.23)$ otherwise. While the small sample size leads to overlapping credible intervals, a significant difference between the two is also supported by a previous experiment that focused on the effects of edge knots \citep{courchene1998effect}. As suggested by the simulation results in Table \ref{tab:sim_parameters}, a real data analysis of 360 to 720 specimens would likely reveal statistical significance. In contrast, a study that recorded only the information of the worst knot on each specimen failed to find evidence of knot location having an impact on UTS \citep{wong2016quantifying}.

\section{Conclusions and recommendations}\label{sec:discussion}

\subsubsection*{Conclusions}
As a bench-scale stage of process development, our case study has developed and applied a new approach to modelling the tensile strength of lumber. The novelty of our approach lies in its incorporation of the spatial arrangement of knots extracted from the surface scans of specimens, and the use of ideas and methods from the theory of spatial statistics.  Thus, the model accounts for the locations and strength-reducing impacts of all relevant knots, as derived from reconstructing their 3-D volumes. The usefulness of our approach is demonstrated via simulation studies and the main goal of our study, namely a framework for modelling the strengths of our $113$ DF specimens. Section~\ref{sec:application} showed that even in this small bench-scale experiment, our proposed method has the potential to outperform more simplistic approaches such as basic regression modelling in predictive performance.  Furthermore, any other non-destructive measurements taken on the specimen, such as MOE, could be incorporated in the modelling framework as local or global covariates. 

Thus, we believe the overarching goal of this case study, a proof-of-concept for our strength prediction approach, has been achieved. Its scientific value lies in showing how different knots on a lumber specimen work together to reduce its strength. Furthermore, the metrics for predictive performance presented in Section~\ref{subsec:model_pred_performance} showed that our Bayesian model achieved MSPEs, MAPEs, and $95\%$ prediction interval lengths that were favourable compared with two other regression models. All three models included MOE, which is recognized to be a strong predictor of UTS and other strength properties of lumber, for example as used in monitoring programs \citep{kretschmann1999monitoring}.  Adding the effect of the worst knot improved predictions, and using the complete spatial arrangement of knots in our Bayesian model yielded further improvements, albeit modest. Our relatively minor gains may be due in part to limitations of the available sample of specimens for this study. The Bayesian model is more complex relative to the amount of information contained in this small sample of lumber, and we would expect to see more substantial gains in samples of the size more commonly seen in lumber testing of over $300$  pieces (see, for example, \citet{astm2007standard}, Note 7). Also, the high grade quality of the DF specimens is less ideal for studying the effects of knots, as the number of knots per specimen was relatively low (average 3.8) and limited the information available for learning knot parameters.

\subsubsection*{Recommendations}

The specimens analyzed in this case study are a sample that represents the DF lumber population of grade L3, with nominal size 2 by 6 and length 12-ft. Thus, the specific model developed and parameter estimates can only be generalized to that specific population. By leveraging engineering models for the size effects of lumber \citep{madsen1986size}, some aspects of strength and knot effects could be extrapolated to other dimensional sizes of DF. But most importantly, it is the approach to developing such Bayesian strength models that is generalizable, where the specific model adaptations will depend on the requirements of lumber manufacturers.

As practical implications of this study, our proof-of-concept shows that more extensive testing and efficient scanning techniques would be desirable to produce larger samples representing more diverse lumber populations (e.g., different species and dimensional sizes). In particular, an interesting extension would be to take multiple strength measurements on a single specimen, to better understand how strength varies along a piece of lumber. We note that even modest gains in strength prediction accuracy would be worthwhile, with the large volumes of lumber being produced worldwide and the increasing popularity of wood as a sustainable building material. More accurate predictions could in turn reduce the within-grade strength variation, via updates to the rules used to classify lumber into grades. This would be beneficial for both manufacturers and consumers of lumber, in terms of extracting better value from harvested trees and meeting strength requirements for intended uses with greater confidence. Overall, the model and approach presented could lead to a more refined way of managing the quality of manufactured lumber. Beyond individual specimens, another application of this modelling approach is in estimating the strength of larger engineered wood products (or ``mass timber'') where lumber is the primary feedstock~\citep[see][for more information]{yard2017mass}.  An example of this is glued laminated timber (glulam) where individual pieces of lumber are glued together so that they resist the applied load in a parallel fashion. The higher load carrying capacities of glulam allows wood to be used in taller and larger buildings. In glulam, it is unlikely that the weak cell of one lamination aligns with the weak cell of the adjacent lamination~\citep{issa2005advanced}. The ability to estimate the strength and variability of glulam without extensive full-size testing of all sizes and grades of large glulam would be an important step towards establishing design values for this product.

\section*{Acknowledgements}

The work reported in this case study was partially funded by FPInnovations and grants from the Natural Science and Engineering Research Council of Canada. The authors thank Conroy Lum, along with FPInnovations and its technical support staff, for facilitating the experimental work that was done to produce the data used in this case study. Thanks also to Conroy for his expertise and assistance in helping the authors understand better the complexities involved in grading lumber and providing constructive comments on the manuscript. Finally, the work profited from the comments of members of the Forest Products Stochastic Modelling Group, centered at the University of British Columbia.

\section*{Data availability statement}
The code and data that support the results of this study are available at the repository \url{https://github.com/wongswk/tensile-strength-model}.

\bibliographystyle{apalike}

\begin{thebibliography}{}
	
	\bibitem[{ASTM International}, 2007]{astm2007standard}
	{ASTM International} (2007).
	\newblock {\em {D1990-07: Standard practice for establishing allowable
			properties for visually-graded dimension lumber from in-grade tests of
			full-size specimens}}.
	\newblock ASTM International, West Conshohocken, Pennsylvania.
	
	\bibitem[{ASTM International}, 2012]{astmvib}
	{ASTM International} (2012).
	\newblock {\em {D6874-12: Standard Test Methods for Nondestructive Evaluation
			of the Stiffness of Wood and Wood-Based Materials Using Transverse Vibration
			or Stress Wave Propagation}}.
	\newblock ASTM International, West Conshohocken, Pennsylvania.
	
	\bibitem[{ASTM International}, 2015]{astmd198}
	{ASTM International} (2015).
	\newblock {\em {D198-15: Standard Test Methods of Static Tests of Lumber in
			Structural Sizes}}.
	\newblock ASTM International, West Conshohocken, Pennsylvania.
	
	\bibitem[{ASTM International}, 2019]{astmtensile}
	{ASTM International} (2019).
	\newblock {\em {D4761-19: Standard Test Methods for Mechanical Properties of
			Lumber and Wood-Base Structural Material}}.
	\newblock ASTM International, West Conshohocken, Pennsylvania.
	
	\bibitem[Barbosa et~al., 2019]{barbosa2019effect}
	Barbosa, M.~C., Street, J., Owens, F.~C., and Shmulsky, R. (2019).
	\newblock The effect of multiple knots in close proximity on southern pine
	lumber properties.
	\newblock {\em Forest Products Journal}, 69(4):278--282.
	
	\bibitem[Battaini et~al., 2000]{battaini2000bench}
	Battaini, M., Yang, G., and Spencer, B. (2000).
	\newblock Bench-scale experiment for structural control.
	\newblock {\em Journal of Engineering Mechanics}, 126(2):140--148.
	
	\bibitem[Biggeri et~al., 1996]{biggeri1996air}
	Biggeri, A., Barbone, F., Lagazio, C., Bovenzi, M., and Stanta, G. (1996).
	\newblock Air pollution and lung cancer in trieste, italy: spatial analysis of
	risk as a function of distance from sources.
	\newblock {\em Environmental health perspectives}, 104(7):750--754.
	
	\bibitem[Bithell, 1995]{bithell1995choice}
	Bithell, J. (1995).
	\newblock The choice of test for detecting raised disease risk near a point
	source.
	\newblock {\em Statistics in Medicine}, 14(21-22):2309--2322.
	
	\bibitem[{Canadian Lumber Standards Accreditation Board}, 2021]{lumbergrades}
	{Canadian Lumber Standards Accreditation Board} (2021).
	\newblock Lumber grades.
	\newblock
	https://www.clsab.ca/faq-items/how-were-the-design-values-established-for-structural-lumber/.
	\newblock Accessed 02/26/2022.
	
	\bibitem[Cast{\'e}ra et~al., 1996]{castera1996prevision}
	Cast{\'e}ra, P., Faye, C., and El~Ouadrani, A. (1996).
	\newblock Prevision of the bending strength of timber with a multivariate
	statistical approach.
	\newblock {\em Annals of Forest Science}, 53(4):885--896.
	
	\bibitem[Courchene et~al., 1998]{courchene1998effect}
	Courchene, T., Lam, F., and Barrett, J. (1998).
	\newblock {The effect of edge knots on the strength of SPF MSR lumber}.
	\newblock {\em Forest products journal}, 48(4):75--81.
	
	\bibitem[Cramer and Goodman, 1983]{cramer2007model}
	Cramer, S. and Goodman, J. (1983).
	\newblock Model for stress analysis and strength prediction of lumber.
	\newblock {\em Wood and Fiber Science}, 15(4):338--349.
	
	\bibitem[Diggle, 1990]{diggle1990point}
	Diggle, P.~J. (1990).
	\newblock A point process modelling approach to raised incidence of a rare
	phenomenon in the vicinity of a prespecified point.
	\newblock {\em Journal of the Royal Statistical Society: Series A (Statistics
		in Society)}, 153(3):349--362.
	
	\bibitem[Divos and Tanaka, 1997]{divos1997lumber}
	Divos, F. and Tanaka, T. (1997).
	\newblock Lumber strength estimation by multiple regression.
	\newblock {\em Holzforschung}, 51(5):467--471.
	
	\bibitem[Fink and Kohler, 2014]{fink2014model}
	Fink, G. and Kohler, J. (2014).
	\newblock Model for the prediction of the tensile strength and tensile
	stiffness of knot clusters within structural timber.
	\newblock {\em European Journal of Wood and Wood Products}, 72(3):331--341.
	
	\bibitem[Foley, 2003]{foley2003modeling}
	Foley, C. (2003).
	\newblock {\em Modeling the effects of knots in structural timber}.
	\newblock PhD dissertation, Lund Institute of Technology.
	
	\bibitem[Fran{\c{c}}a et~al., 2018]{francca2018modeling}
	Fran{\c{c}}a, F., Seale, R.~D., Shmulsky, R., and Fran{\c{c}}a, T. (2018).
	\newblock Modeling mechanical properties of 2 by 4 and 2 by 6 southern pine
	lumber using longitudinal vibration and visual characteristics.
	\newblock {\em Forest Products Journal}, 68(3):286--294.
	
	\bibitem[Garc{\'\i}a and Rosales, 2017]{garcia2017deflections}
	Garc{\'\i}a, D.~A. and Rosales, M.~B. (2017).
	\newblock Deflections in sawn timber beams with stochastic properties.
	\newblock {\em European Journal of Wood and Wood Products}, 75(5):683--699.
	
	\bibitem[Gelman et~al., 2013]{gelman2013bayesian}
	Gelman, A., Carlin, J.~B., Stern, H.~S., Dunson, D.~B., Vehtari, A., and Rubin,
	D.~B. (2013).
	\newblock {\em Bayesian Data Analysis}.
	\newblock Chapman and Hall/CRC.
	
	\bibitem[Gelman et~al., 2017]{gelman2017prior}
	Gelman, A., Simpson, D., and Betancourt, M. (2017).
	\newblock The prior can often only be understood in the context of the
	likelihood.
	\newblock {\em Entropy}, 19(10):555.
	
	\bibitem[Green and McDonald, 1993]{green1993investigation}
	Green, D.~W. and McDonald, K.~A. (1993).
	\newblock Investigation of the mechanical properties of red oak 2 by 4's.
	\newblock {\em Wood and fiber science}, 25(1):35--45.
	
	\bibitem[Harte, 2017]{harte2017mass}
	Harte, A.~M. (2017).
	\newblock Mass timber -- the emergence of a modern construction material.
	\newblock {\em Journal of Structural Integrity and Maintenance}, 2(3):121--132.
	
	\bibitem[Hietaniemi and Silv{\'e}n, 2011]{hietaniemi2011camera}
	Hietaniemi, R. and Silv{\'e}n, O. (2011).
	\newblock Camera based lumber strength classification system.
	\newblock In {\em MVA2011 IAPR Conference on Machine Vision Applications},
	pages 251--254.
	
	\bibitem[Issa and Kmeid, 2005]{issa2005advanced}
	Issa, C.~A. and Kmeid, Z. (2005).
	\newblock Advanced wood engineering: glulam beams.
	\newblock {\em Construction and Building Materials}, 19(2):99--106.
	
	\bibitem[Jun et~al., 2019]{jun2019sequential}
	Jun, S.-H., Wong, S.~W., Zidek, J.~V., and Bouchard-C{\^o}t{\'e}, A. (2019).
	\newblock Sequential decision model for inference and prediction on nonuniform
	hypergraphs with application to knot matching from computational forestry.
	\newblock {\em The Annals of Applied Statistics}, 13(3):1678--1707.
	
	\bibitem[Kandler et~al., 2015]{kandler2015stochastic}
	Kandler, G., F{\"u}ssl, J., and Eberhardsteiner, J. (2015).
	\newblock Stochastic finite element approaches for wood-based products:
	theoretical framework and review of methods.
	\newblock {\em Wood science and technology}, 49(5):1055--1097.
	
	\bibitem[Kretschmann et~al., 1999]{kretschmann1999monitoring}
	Kretschmann, D.~E., Evans, J.~W., and Brown, L. (1999).
	\newblock {\em Monitoring of visually graded structural lumber}, volume 576.
	\newblock US Department of Agriculture, Forest Service, Forest Products
	Laboratory.
	
	\bibitem[Lam and Varoglu, 1991]{lam1991variation}
	Lam, F. and Varoglu, E. (1991).
	\newblock Variation of tensile strength along the length of lumber.
	\newblock {\em Wood science and technology}, 25(5):351--359.
	
	\bibitem[Lukacevic et~al., 2015]{lukacevic2015discussion}
	Lukacevic, M., F{\"u}ssl, J., and Eberhardsteiner, J. (2015).
	\newblock Discussion of common and new indicating properties for the strength
	grading of wooden boards.
	\newblock {\em Wood Science and Technology}, 49(3):551--576.
	
	\bibitem[Madsen and Buchanan, 1986]{madsen1986size}
	Madsen, B. and Buchanan, A.~H. (1986).
	\newblock Size effects in timber explained by a modified weakest link theory.
	\newblock {\em Canadian Journal of Civil Engineering}, 13(2):218--232.
	
	\bibitem[NLGA, 2017]{nlga2017}
	NLGA (2017).
	\newblock {\em Standard Grading Rules for Canadian Lumber}.
	\newblock National Lumber Grades Authority, Vancouver, Canada.
	
	\bibitem[Oh et~al., 2010]{oh2010use}
	Oh, J.-K., Kim, K.-M., and Lee, J.-J. (2010).
	\newblock Use of adjacent knot data in predicting bending strength of dimension
	lumber by x-ray.
	\newblock {\em Wood and fiber science}, 42(1):10--20.
	
	\bibitem[Pan et~al., 2021]{pan2021ellipse}
	Pan, S., Fan, S., Wong, S.~W., Zidek, J.~V., and Rhodin, H. (2021).
	\newblock Ellipse detection and localization with applications to knots in sawn
	lumber images.
	\newblock In {\em Proceedings of the IEEE/CVF Winter Conference on Applications
		of Computer Vision}, pages 3892--3901.
	
	\bibitem[Reynolds and Pierson, 1923]{reynolds1923lumber}
	Reynolds, R. V.~R. and Pierson, A.~H. (1923).
	\newblock {\em Lumber Cut of the United States, 1870-1920}.
	\newblock Number 1119. US Department of Agriculture.
	
	\bibitem[{Stan Development Team}, 2020]{stan2018rstan}
	{Stan Development Team} (2020).
	\newblock {RStan: the R interface to Stan}.
	\newblock {\em R package version 2.21.2.}
	
	\bibitem[Taylor et~al., 1992]{taylor1992comparing}
	Taylor, S., Bender, D., Kline, D., and Kline, K. (1992).
	\newblock Comparing length effect models for lumber tensile strength.
	\newblock {\em Forest products journal}, 42(2):23--30.
	
	\bibitem[Tegel et~al., 2012]{tegel2012early}
	Tegel, W., Elburg, R., Hakelberg, D., St{\"a}uble, H., and B{\"u}ntgen, U.
	(2012).
	\newblock Early neolithic water wells reveal the world's oldest wood
	architecture.
	\newblock {\em {PloS ONE}}, 7(12):e51374.
	
	\bibitem[Vikram et~al., 2011]{vikram2011stiffness}
	Vikram, V., Cherry, M.~L., Briggs, D., Cress, D.~W., Evans, R., and Howe, G.~T.
	(2011).
	\newblock Stiffness of Douglas-Fir lumber: effects of wood properties and
	genetics.
	\newblock {\em Canadian Journal of Forest Research}, 41(6):1160--1173.
	
	\bibitem[Wong et~al., 2016]{wong2016quantifying}
	Wong, S.~W., Lum, C., Wu, L., and Zidek, J.~V. (2016).
	\newblock {Quantifying uncertainty in lumber grading and strength prediction: a
		Bayesian approach}.
	\newblock {\em Technometrics}, 58(2):236--243.
	
	\bibitem[Yard, 2017]{yard2017mass}
	Yard, A. (2017).
	\newblock Mass timber in North America.
	\newblock {\em ARCHITECT}, 106(5):54--57.
	
	\bibitem[Zidek and Lum, 2018]{zidek2018statistical}
	Zidek, J.~V. and Lum, C. (2018).
	\newblock Statistical challenges in assessing the engineering properties of
	forest products.
	\newblock {\em Annual Review of Statistics and Its Application}, 5:237--264.
	
\end{thebibliography}

\end{document}